# Investigating Streaming Techniques and Energy Efficiency of Mobile Video Services

Mohammad Ashraful Hoque*, Matti Siekkinen*, Jukka K. Nurminen*, Mika Aalto†
*Aalto University School of Science
Email: {mohammad.hoque,matti.siekkinen,jukka.k.nurminen}@aalto.fi
†Nokia Siemens Network
Email: mika.aalto@nsn.com

***Abstract*—We report results from a measurement study of three video streaming services, YouTube, Dailymotion and Vimeo on six different smartphones. We measure and analyze the traffic and energy consumption when streaming different quality videos over Wi-Fi and 3G. We identify five different techniques to deliver the video and show that the use of a particular technique depends on the device, player, quality, and service. The energy consumption varies dramatically between devices, services, and video qualities depending on the streaming technique used. As a consequence, we come up with suggestions on how to improve the energy efficiency of mobile video streaming services.**

## I. Introduction

Digital video content is increasingly consumed in mobile devices [1]. For example, in 2011 YouTube had more than 1 trillion global views and 10% of it was accessed via mobile phones or tablets [2]. At the same time battery life of smartphones has become a critical factor in user satisfaction [3]. Therefore, it is essential that mobile video streaming not only provides a good viewing experience but also avoids excessive energy consumption.

In this paper, we study three popular video streaming services, namely YouTube, Dailymotion and Vimeo, on smartphones. We focus specifically on the video delivery method, which we refer to as a streaming technique, and the resulting energy consumption characteristics. Although the look and feel of different video services may be similar, we have identified six different streaming techniques. Some download the entire video at once, while others may receive it in large chunks. The technique chosen depends on the particular service, the client device, and video resolution.

At present it is not well understood how the different techniques are chosen and how they compare to each other and what are the optimal techniques to use in different contexts. For this reason, we measured the traffic and energy consumption during a large number of streaming sessions from these three services using six different smartphones, covering all the major mobile platforms, and two access network technologies, namely Wi-Fi and 3G.

Our results reveal that different smartphones apply different methods in different contexts. There is little consensus; different techniques are used by different clients to access the same service in the same context. As for the energy consumption, the results vary dramatically between the different devices and streaming techniques, even between seemingly similar techniques. To the best of our knowledge, we are the first to uncover the way the popular streaming services deliver video to different smartphones and to quantify the resulting energy consumption for both Wi-Fi and 3G. We believe that our findings are key to optimizing the energy consumption of mobile video streaming services.

The main contributions of this paper are:

1) We identify and classify the different streaming techniques used by three widely popular video streaming services. We identify the attributes on which the strategy selection depends (client device, player type, content quality, video provider).
2) We captured the streaming traffic and energy consumption of more than 500 streaming sessions and use the measurement results to analyze how different techniques influence the energy consumption of a mobile device.
3) We discuss the implications of our findings and outline strategies on how the streaming services could be further optimized for energy efficiency.

We structure our paper as follows. In the next section, we briefly describe the energy consumption characteristics of wireless communication on smartphones and explain how the common mobile streaming services work. In Section III, we describe our methodology. In Section IV, we explain the different streaming techniques and investigate which services use which ones. Section V is devoted to presenting the results from the energy consumption measurements. Finally, we contrast our work with earlier research in section VI before concluding the paper.

## II. Background

Smartphones allow users to access Internet via Wi-Fi and mobile broadband interface. Currently WCDMA/HSPA is the most widely deployed mobile broadband interface in mobile phones. LTE is another mobile broadband technology commonly available in certain markets. In this paper we focus on WCDMA/HSPA and referred to it as 3G. The power consumption of these interfaces can be very high even though there are already existing standard mechanisms for power saving. The power saving mechanisms of the radio layers use protocols and state changes independently without any knowledge about the applications being used. In this section we briefly review the power consumption characteristics of these two main network interfaces that we use in this study. Then, we explain the typical characteristics of a mobile streaming service.

### *A. Power Saving Mechanisms for Wi-Fi & 3G*

Wi-Fi interface of a smartphone can operate in power saving mode (802.11 PSM). It is a cooperative mechanism and works as follows. PSM allows the Wi-Fi radio periodically to switch into sleep state. The client device wakes up after every 100 ms to receive traffic indication map (TIM) message, which is also known as beacon, from the access point (AP). The received beacon contains information whether the AP has buffered data for the client or not. If there is any data buffered, the client sends PS-Poll frame to the AP. Otherwise the client goes back to sleep until the next beacon. Modern Wi-Fi-enabled devices usually implement a timer which keeps the interface in idle state for a (few) hundred milliseconds after the transmission or reception packets, which improves the performance of short TCP connections.

If PSM is not used, the Wi-Fi interface remains in continuous active mode (CAM). Receiving constant bit rate multimedia streaming traffic using PSM can sometimes resemble CAM. The reason is that constantly arriving packets do not leave enough time in between receiving packets for the timer to expire in order to enter sleep mode. However, if the multimedia traffic is bursty in nature, i.e. multiple packets are sent together as a burst, and the intervals between the bursts are long enough, PSM can save energy.

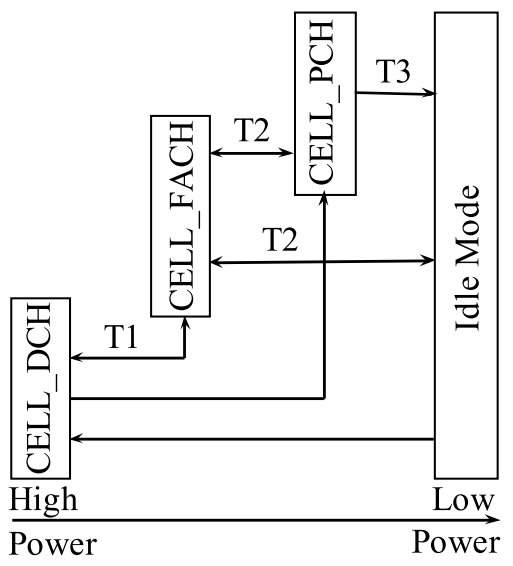

**Fig. 1:** 3G RRC state machine with CELL_DCH, CELL_FACH, CELL_PCH states.

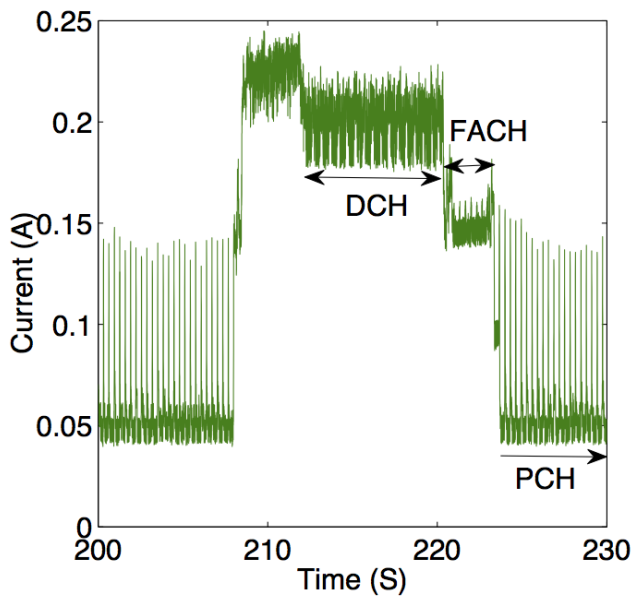

**Fig. 2:** Current consumption at different states and state transitions with Lumia 800.

In case of cellular networks, such as 3G, the standards specify how to control radio resources in such manner that mobility in cellular network as well mobile device power consumption can be optimized. From power consumption point of view, the radio resource control and states and transitions between states must be understood. Figure 1 shows the states and the inactivity timers in 3GPP RRC protocol. These timers are used by the 3G radio network to control the transitions among different states and network configuration of these timers has impact on radio resource usage, power consumption and user experience.

Figure 2 shows that average current consumption in CELL_DCH state is 200 mA, in CELL_FACH state 150 mA, and CELL_PCH 50mA, approximately. However, the timers T1, T2 and T3 have static values and there is no standard procedure on deciding their values. Network operators use different configurations in the radio network depending also on the network equipment capabilities provided by the network vendors. Inactivity timer settings in live networks vary from few seconds up to tens of seconds. The potential consequence especially with long inactivity timers is high power consumption at the mobile device.

### *B. Mobile Streaming*

Mobile streaming services today deliver data most commonly using HTTP over TCP, similarly to non-mobile streaming. Mobile device users can have two ways of accessing the service, by using either a native app or browser which loads a Flash or HTML5 player in the beginning of the streaming session.

A common feature for all streaming services is an initial buffering of multimedia content at the client which tries to ensure smooth playback in the presence of bandwidth fluctuation and jitter. This buffering is visible to user as start-up delay and we refer to this phase as Fast Start. The name comes from the fact that this initially buffered data is typically downloaded at full speed, i.e. by using all the available bandwidth, while the rest of the video necessarily is not.

The quality of the video played is often denoted with a p-notation, such as 240p, which refers to the resolution of the video. 240p usually refers to 360x240 resolution. Different services use also low, standard, and high definition (LD, SD, HD) notations but the resolutions that each one refers to varies between services. Therefore, we define 240p videos as LD, 270-480p videos as SD and 720-1080p videos as HD.

## III. Methodology

We used three popular streaming services, namely YouTube, Vimeo and Dailymotion, and six different smartphones which cover all the major mobile platforms. Some of the services have a native app for some mobile platforms. For example, YouTube apps exist for all five platforms, whereas Symbian and Meego apps do not exist for Vimeo and Dailymotion. However, it is possible to watch Vimeo videos using a browser on Nokia 701 (Symbian) and N9 (Meego). In Android phones, Nexus S and Galaxy S3, we used both app and browser to access YouTube and Dailymotion videos. Whenever available for the particular smartphone and player, we streamed different quality videos, namely LD, SD, and HD, which range from 240p to 1080p.

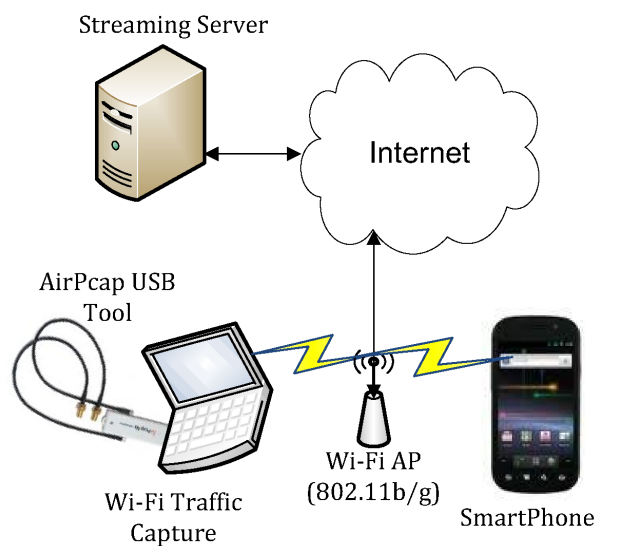

**Fig. 3:** Collecting video traffic from Wi-Fi network.

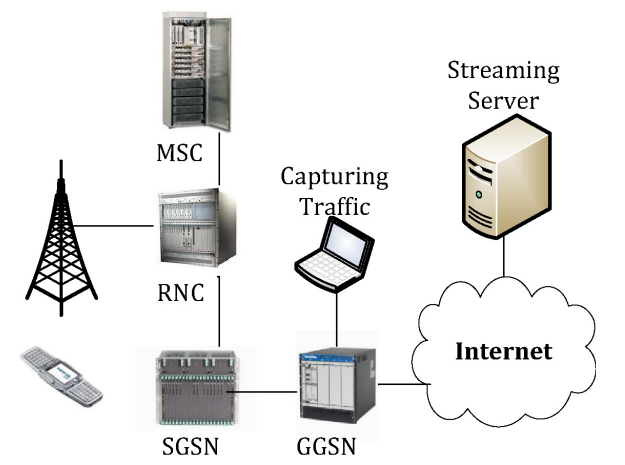

**Fig. 4:** Collecting video traffic from 3G network.

We streamed the videos of our target video services to the different devices via both Wi-Fi and 3G. In total, we captured the streaming traffic of more than 500 streaming sessions. Capturing Internet traffic in a laptop or desktop machine is trivial. Existing tools such as tcpdump or wireshark can be used in the computer. However, such tools are rarely available for mobile platforms. Therefore, we used a traffic capturing tool called AirPcap [4] in a windows machine in monitoring mode. We used a DLink DIR-300 wireless AP supporting 802.11 b/g (54Mbps), which was connected to the Internet via 100 Mbps Ethernet. The setup is sketched in Figure 3.

We performed 3G measurements in a completely isolated RF room where we had access to a complete HSPA test network provided by Nokia Siemens Networks. The test network was configured according to the vendor recommended parameters The most relevant parameters for this study are the inactivity timers (T1 = 8 s, T2 = 3 s and T3 = 29 mins) and CELL_PCH state being enabled. We captured traffic of the streaming clients from the Gn interface i.e. between SGSN and GGSN as shown in Figure 4. The downlink capacity of the 3G subscription was 6 Mbps.

In order to understand the energy consumption characteristics, we measured the current draw of the smartphones during the streaming sessions. We used two instruments: Monsoon Power Monitor [5] and another custom power monitor. One of these was attached to the phone to measure the current consumption and voltage during a complete video playing period. We detached the phone batteries and powered the phones directly from the power monitor (Monsoon) or using an external power supply.

| Service | Properties | Nokia-N9 (Meego) | Nokia-701 (Symbian) | iPhone-4 (iOS 5.0) | Nexus S(Android-2.3.6) | | Galaxy S3(Android-4.0.4) | | Lumia-800 (WP 7.5) |
|---|---|---|---|---|---|---|---|---|---|
| | | | | | browser | app | browser | app | |
| **YouTube** | Streaming via Wi-Fi& 3G | Encoding rate | Encoding rate | Throttling Factor=2.0 | Throttling Factor=1.25 | On-Off | Encoding rate(HD), Throttling(<HD) Factor=1.25 | On-Off | Fast Caching |
| | Video Quality | LD(240p), SD(270p) | LD(240p), SD(270p) | LD(240p), SD(270p), HD(720p) | LD(240p), SD(360p) | LD(240p), SD(360,480p) | LD(240p), SD(360,480p), HD(720,1080p) | LD(240p), SD(360,480p), HD(720p) | LD(240p), SD(360,480p) |
| | Video Container | mp4(270p) 3gpp(240p) | mp4(270p) 3gpp(240p) | mp4(360,720p) 3gpp(270p) | xflv | mp4(360p) 3gpp(240p) | xflv | mp4(>240p) 3gpp(240p) | mp4(270p) 3gpp(240p) |
| **Vimeo** | Streaming via Wi-Fi& 3G | Encoding rate | Fast Caching | DASH | On-Off | | On-Off | | Fast Caching |
| | Video Quality | SD(270p) | SD(270p) | SD(270,480p), HD(720p) | SD(270p) | | SD(270p) | | SD(270p) |
| | Video Container | mp4 | mp4 | mp4 | mp4 | | mp4 | | mp4 |
| **Daily-motion** | Streaming via Wi-Fi& 3G | Throttling Factor=1.25 | – | Throttling Factor=1.25 | On-Off | | Fast Caching(288p), On-Off(>288p) | | Throttling Factor=1.25 |
| | Video Quality | SD(288p) | – | LD(240), SD(288,480p) | SD(270p) | | SD(288,480p),HD(720p) | | SD(288p) |
| | Video Container | mp4 | – | mp4 | mp4 | | mp4 | | mp4 |

**TABLE I:** Streaming techniques for popular video streaming services to mobile phones of five platforms. Streaming technique does not depend on the wireless interface being used for, rather depends on the player, video quality, device and the video service provider.

## IV. Video Streaming Services and Techniques

We show in this section that the way the content is delivered to the client after the initial buffering, which we call streaming technique, heavily depends on the service and client characteristics. We have identified five different techniques that the three investigated services used. We explain each technique below.

First, we need to clarify the interplay between TCP and the streaming server and client applications. Consider the simplified illustration in Figure 5. When the server wants to send content to the client player, it stores that data to TCP send buffer via socket API. TCP will transmit the data if there is room in congestion and receive window. The player reads from the TCP receive buffer through the socket API and stores the data into a playback buffer. If the player pauses reading from the receive buffer, it becomes full and TCP flow control activates by sending zero window advertisements. Then, sending TCP must pause transmission. In this way, the behavior of the player can determine the way the data is delivered from the server. On the other hand, if the client player reads constantly from TCP socket, the server is in control of the transmission rate. Finally, if also the server sends data without rate limitation, then the amount of available bandwidth on the TCP/IP path determines the transmission rate.

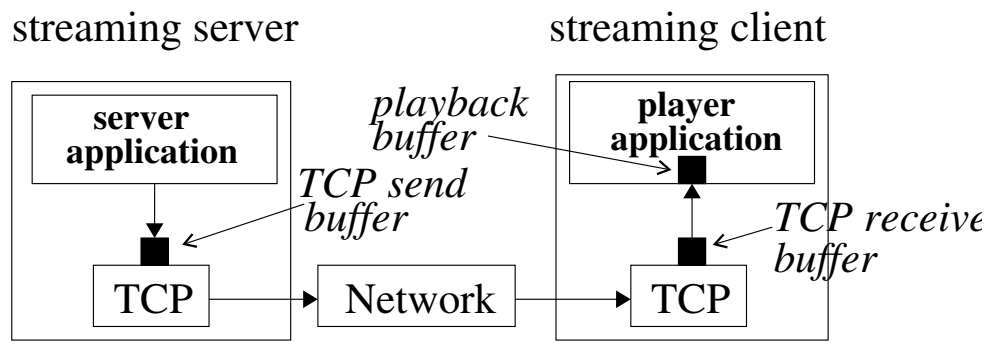

**Fig. 5:** Interplay of TCP and applications via buffers.

**(i) Encoding-Rate streaming:** Stream data is delivered at the encoding rate. The client player reads data from TCP socket at the rate of consumption which is equal to the encoding rate. The server would send data at a higher rate but receiving TCP's buffer fills up and TCP flow control prevents it.

**(ii) Server Throttling:** The content is received by the client at a constant rate but higher than the encoding rate. In this case it is the server that controls the transmission rate and client reads data from TCP buffer continuously. Consequently, the player keeps on accumulating data in the playback buffer throughout the streaming session.

**(iii) On-off streaming:** The stream data is delivered in the form of bursts in between which virtually no data is delivered. The client player reads from the TCP buffer periodically in which case TCP flow control pauses the transmission after the buffer has filled up again.

**(iv) Fast Caching:** The transmission rate is limited neither by the server nor by the client but only by the available bandwidth on the path between the server and client. In other words, server sends stream data at full rate and client reads it from TCP buffer continuously.

**(v) Dynamic Adaptive Streaming over HTTP (DASH):** DASH is a fundamentally different streaming technique. In all the previous techniques, the video quality is fixed in the beginning and cannot be changed in the middle of a streaming session. In DASH, the player can dynamically switch between qualities. The server maintains several copies of the same video of different quality, each one broken into segments typically worth several seconds of video. The client player requests each video segment separately and chooses the quality of each segment to request depending on the available bandwidth and its fluctuations. The transmission of all the segments take place either using a single TCP connection or via multiple TCP connections.

We inferred manually the type of streaming technique used for each of the different combinations of device, service, stream quality, player type (app or browser), and access network type (Wi-Fi or 3G). Table I summarizes the results. It shows that there is no systematic use of a given technique by a given streaming service. Neither is a particular technique tied to specific mobile platforms. Instead, the technique used depends on their combination plus the stream quality and the player type. However, the wireless interface being used does not seem to influence the choice of technique.

Next, we describe each technique in detail with illustrations computed from example traffic traces. Although we analyzed all the cases in Table I, we have room to visualize only a few examples.

### A. Encoding-Rate Streaming

YouTube players in N9, Nokia 701 and Galaxy S3 and the Dailymotion player in N9 use ENC technique, as noted in Table I.

The session begins with Fast Start and then the player receives content at the encoding rate. We noted that the sending rate during a session is not exactly constant. Small fluctuations are due to variable bit-rate encoding used in the videos streamed in these examples. The player reads from the full TCP receive buffer at the same rate as it consumes the data and TCP flow control mandates the sending TCP to follow this rate. Figure 6 confirms that zero window advertisements are sent frequent by the receiving TCP.

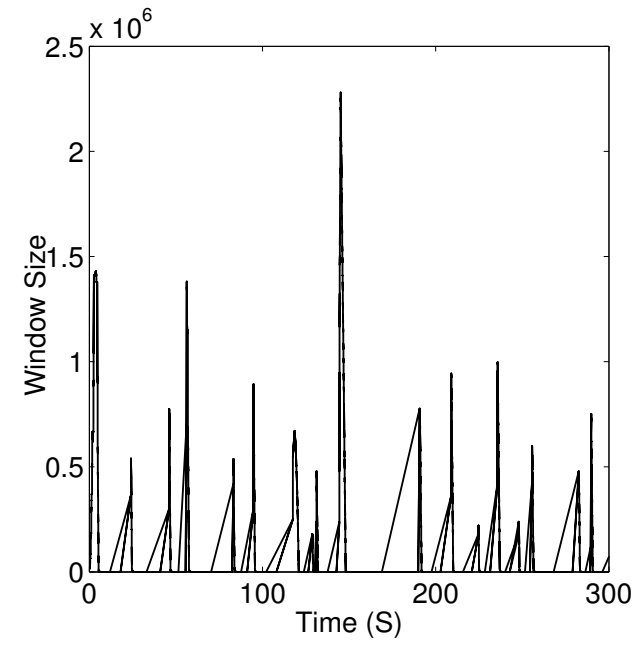

**Fig. 6:** TCP receiver window size (in Bytes) advertised by Nokia N9 alternates between zero and a higher.

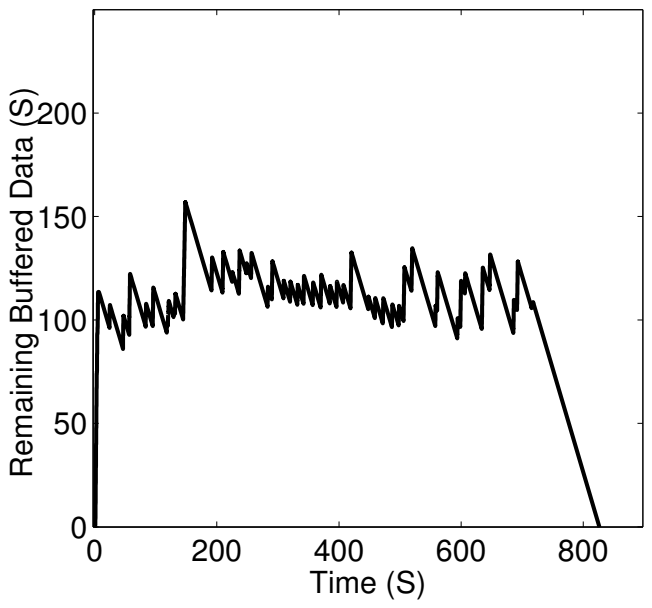

**Fig. 7:** Vimeo player buffers constantly about 100s worth of content while streaming SD video to N9.

We used VLC to watch these particular example videos on a PC (YouTube allows watching specific quality using the `fmt` parameter) and extracted the video encoding rate per second from the player statistics and received data as a function of time from the original traffic traces. With these two time series we computed the estimate of buffered data at the client as a function of time. We show one such graph for N9 Vimeo player in Figure 7 which confirms that the amount of buffered data stays roughly at the same level during the entire streaming session. The amount of data buffered during the Fast Start phase depends on the device and service.

### B. Throttling

Table I indicates that streaming YouTube video to iPhone, Nexus S and Galaxy S3, and streaming Dailymotion videos to N9, iPhone and Lumia 800 use the throttling technique. However, the technique is used only for streaming LD($240p$) and SD($270-480p$) videos and, in the case of Android phones, only when using browser based Flash player. As usual, content is buffered in the beginning during Fast Start phase for a few seconds. Afterwards the server throttles the sending rate and serves the client at a higher than the encoding rate. An example throughput graph in Figures 8 visualizes the behavior.

Again the amount of data buffered during the initial Fast Start phase varies between the players and devices. By comparing the average encoding rate with the amount of bytes received during Fast Start, we estimate the Flash player in mobile browsers of Nexus S and Galaxy S3 to download 40 seconds worth of content from YouTube. Interestingly, the YouTube Flash player specifies this amount as the `burst=40` parameter in the URL for any video request of quality lower than $720p$. However, the YouTube player in iPhone does not use such parameter. We estimate that the server sends it 30 seconds worth of content during Fast Start. In the case of

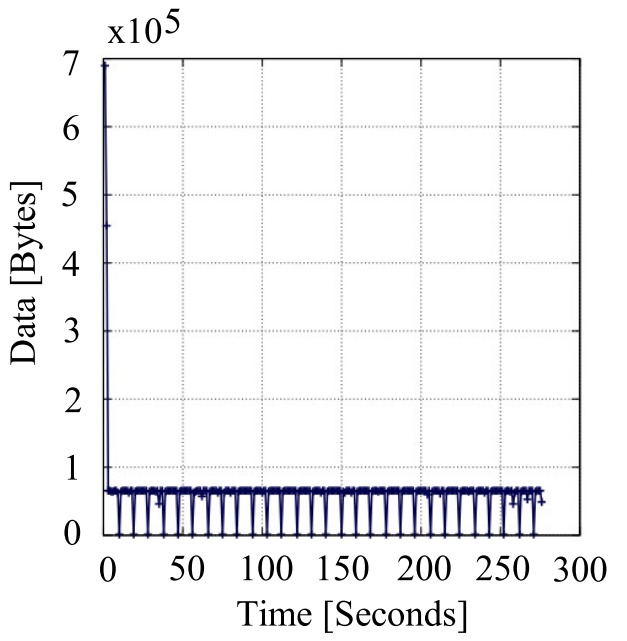

**Fig. 8:** Throughput of a YouTube video streaming to iPhone via 3G with throttle factor 2.0.

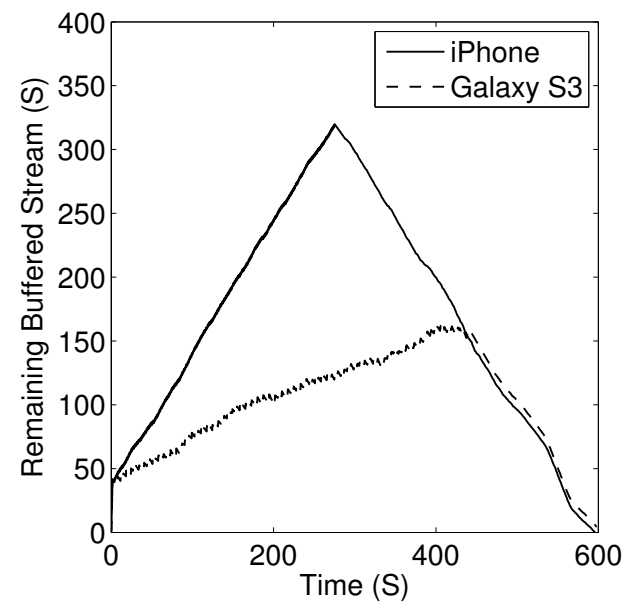

**Fig. 9:** Playback buffer status while playing a YouTube $240p$ video to iPhone and Galaxy S3.

Dailymotion, the amount of initially buffered content is worth 15 seconds for iPhone, N9 and Lumia800.

The rate at which the server sends the stream data also depends on the case. Using the traffic traces, we computed the ratio of throughput and average encoding rate and present it as *throttle factor* in Table I. YouTube servers send video stream at 1.25 times the encoding rate to the Flash player in mobile browsers for LD and SD videos. The player specifies this rate in the URL as the combination of two parameters `algorithm=throttle-factor` and `factor = 1.25`. Although iPhone's YouTube player also receives traffic from the YouTube server in a similar fashion, it does not specify the throttle factor in the URL. The traffic traces reveal that iPhone player receives data at a rate which is roughly twice the encoding rate during the throttling phase. Similar to YouTube, the throttle factor for Dailymotion videos is also 1.25. The Figure 9 exemplifies that the amount of buffered data keeps on accumulating until the download finishes. In addition, we checked that the TCP receiver advertised window is large throughout the streaming sessions which confirms that the rate is indeed controlled by the server and not the client in these cases.

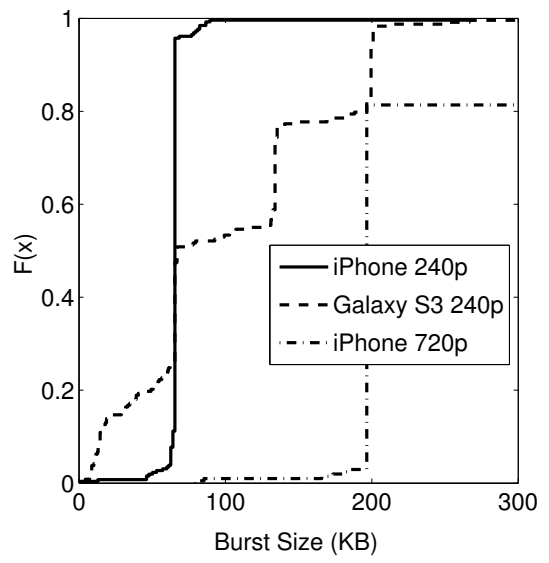

**Fig. 10:** CDF of the burst size for different YouTube streams.

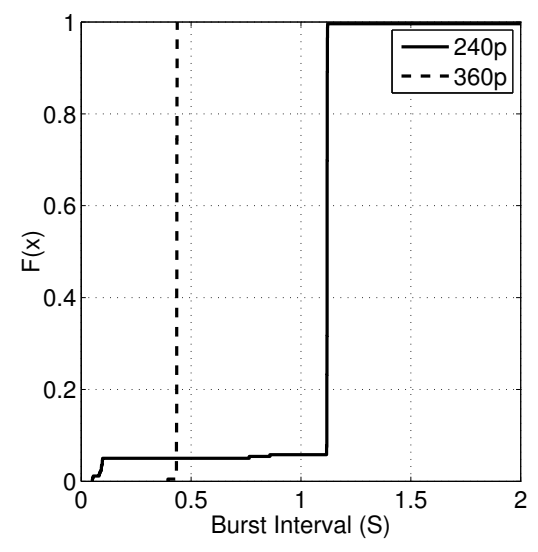

**Fig. 11:** CDF of the burst interval for YouTube streaming to iPhone.

A specific feature of YouTube is that even though the average rate is throttled by the server, it sends the data of LD and SD quality videos to Nexus S, Galaxy S3 and iPhone in a bursty manner. The bursts are separated by a fixed periodic interval. This behavior is similar to the on-off technique (next Section) but the time scale here is relatively short, a few seconds at most in between bursts, while the on-off streaming generates much larger bursts separated by tens of seconds. Nevertheless, also this shorter time scale burstiness has important implications to power consumption, as we show in Section V. Figure 10 shows a CDF plot of the burst sizes computed so that packets with interval shorter than 50ms were grouped together in a burst. We observe, like the authors in [6] and [7], that YouTube servers sends 64KB bursts periodically to a Flash player in mobile browser. However, we also noticed that the burst size for iPhone is 192 KB when streaming video of $720p$ quality. Furthermore, several burst sizes seem to be used in the case of Galaxy S3. Figure 11 plots a CDF of the intervals between first packet of new burst and last packet of the previous burst when streaming to iPhone using two different qualities. We observe that this interval decreases as the encoding rate of the video increases but the burst size remains the same.

*1) Many connections with iPhone:* iPhone app uses multiple TCP connections when streaming HD quality YouTube videos. Video delivery happens in several steps and for each step a separate TCP connection is created sequentially.The player maintains a fixed playback buffer and the player closes the existing TCP connection with the server whenever this buffer is filled. After some buffer space is freed the player again initiates another TCP connection with the server.

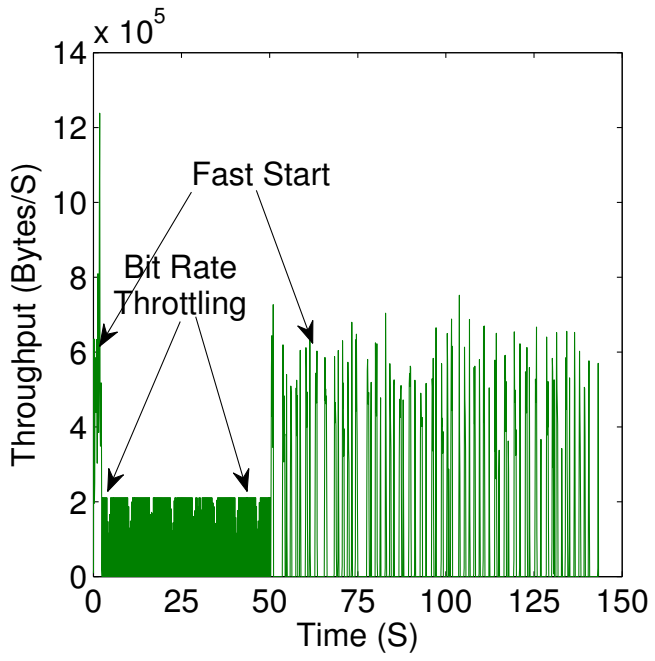

**Fig. 12:** Fast Start and throttling phases while streaming a YouTube HD video to iPhone.

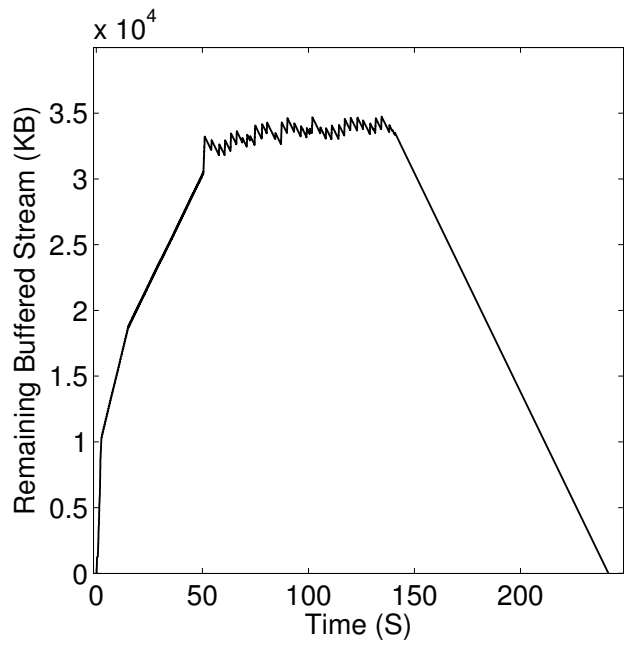

**Fig. 13:** YouTube playback buffer status in iPhone while watching the HD video.

Figure 12 shows the resulting traffic patterns when streaming over Wi-Fi. The entire video of size 75MB is received using 66 connections. The stream begins, as usual, with Fast Start, after which the server throttles the sending rate at twice the encoding rate. This phase happens over a single connection and most of the data is transferred in this phase. For the rest of the requests Fast Start happens again since for each new request the server allows the client to initially buffer data in this way. However, throttle phase is no longer visible as the client player closes the connection with TCP reset packets as soon as the remaining bytes of the fixed playback buffer filled. Figure 13 shows an estimate of the playback buffer status which reveals that the player maintains a fixed playback buffer of about 33 MB.

We noticed, as did ERS et al. [8], that the player receives more data than the actual video size. We computed that during an example streaming session the player receives in total twice the actual playback content, which has a significant impact on users who have, e.g., a monthly quota but also to the network operators given the popularity of YouTube and iPhone. One possible explanation for this behavior is the following: The player does not know the end position of the current key frame or the beginning of the next key frame beforehand and it can close the connection in the middle of a key frame transmission. Consequently, for each new request, the player specifies a range starting from the beginning of a key frame which was received partially during the previous request, and all the data of the previously partially received key frame is wasted.

### C. On-Off

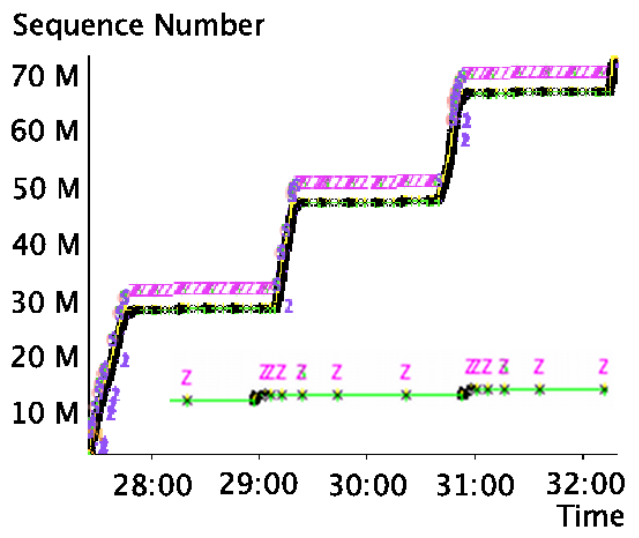

**Fig. 14:** Streaming a 6 min Dailymotion HD video via Wi-Fi to Galaxy S3. Bursts are separated by around 80 seconds. Zoom-in shows zero win advertisements.

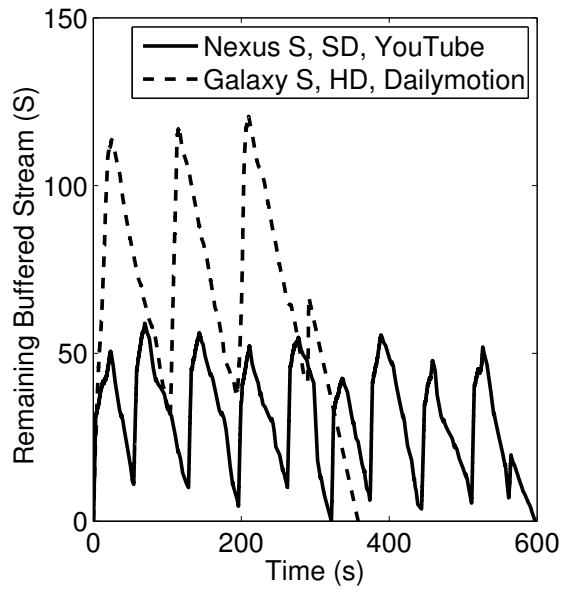

**Fig. 15:** Amount of buffered data during Nexus S YouTube and Galaxy S3 Dailymotion streaming sessions.

*1) Client in control:* On-Off streaming technique namely generates an on-off traffic pattern consisting of large bursts separated by correspondingly long idle intervals. Client player causes this pattern by reading from the TCP socket periodically. As a result, filled up TCP receive buffer and flow control make sure that the TCP sender at the server end must pause the transmission in between these reading events. The behavior is illustrated in Figure 14. The bursts are easy to identify. In between bursts, because the TCP sender at the server side has data from the server application buffered to be sent, it sends zero size packets (zero window probes) in order to check whether the TCP receiver's buffer status has changed. The receiver replies with an ACK with zero window size (Z in the figure). When the player application realizes that the amount of buffered stream drops below a certain threshold, it reads a large burst of data from TCP socket again, which allows TCP to open up the receive window and sender to continue to transmit data. The amount of buffered data through the streaming session is plotted in Figure 15 for two example cases. We observe that Nexus S YouTube app almost completely empties the buffer before reading again from the TCP socket, whereas the Galaxy S3 Dailymotion player keeps more data in the buffer.

*2) Multiple TCP connections with Galaxy S3:* Galaxy S3 app uses multiple TCP connections to stream the video. The streaming mechanism in Galaxy S3 is otherwise the same as the Nexus S except that the YouTube player in Galaxy S3 closes the current TCP connection when the player has enough data in the playback buffer to play. Then, it sends another request when the data falls below a threshold by specifying `Range:` in the HTTP header. Galaxy S3 also downloads some extra bytes but the amount is negligible compared to iPhone (Sec IV-B1). Figure 16 shows the resulting traffic pattern. The amount of buffered data during the streaming session progresses in the same fashion as in Figure 15. Since TCP connection is closed in between bursts, no window advertisements are sent during that time, which has a big impact on the energy consumption, as we show in Section V-C.

### D. Fast Caching

Fast Caching refers to downloading the content immediately using the all available bandwidth between the server and the mobile client. We found that Lumia800 downloads YouTube and Vimeo videos and Nokia 701 Vimeo videos in this way. In the case of YouTube, the player uses `ratebypass=yes` parameter in the HTTP request. The example of Nokia 701 shown in Figure 17 illustrates how the amount of buffered data over time looks like with Fast Caching.

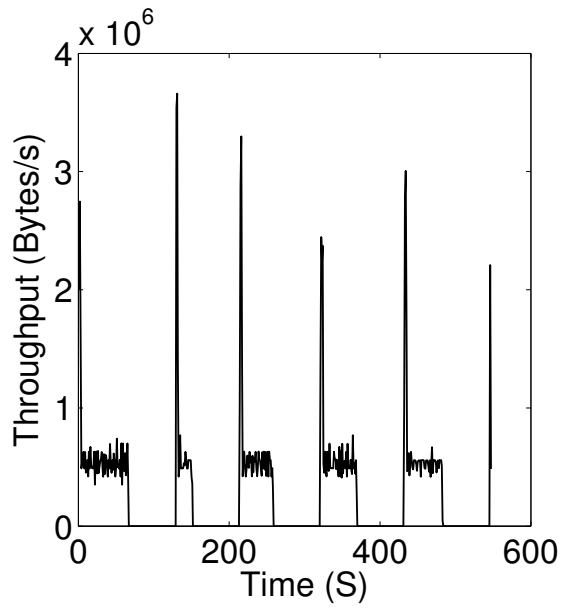

Fig. 16: The YouTube app in Galaxy S3 receives each burst of bytes of a HD video in a separate TCP connection.

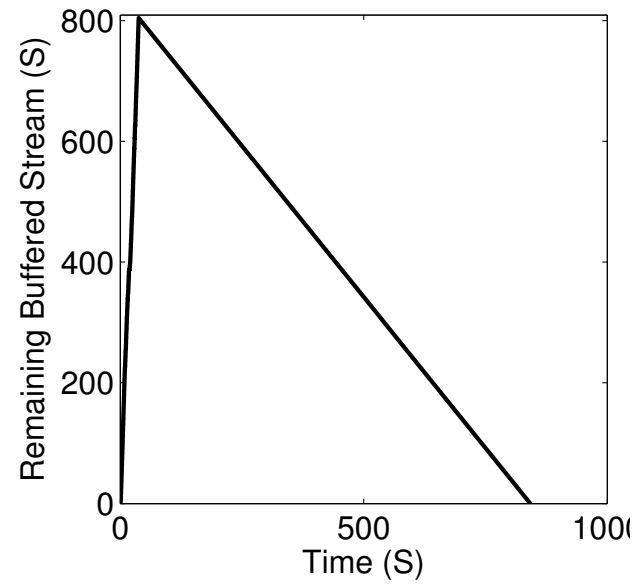

Fig. 17: Amount of buffered data during a Vimeo streaming session using Nokia 701 over 3G network.

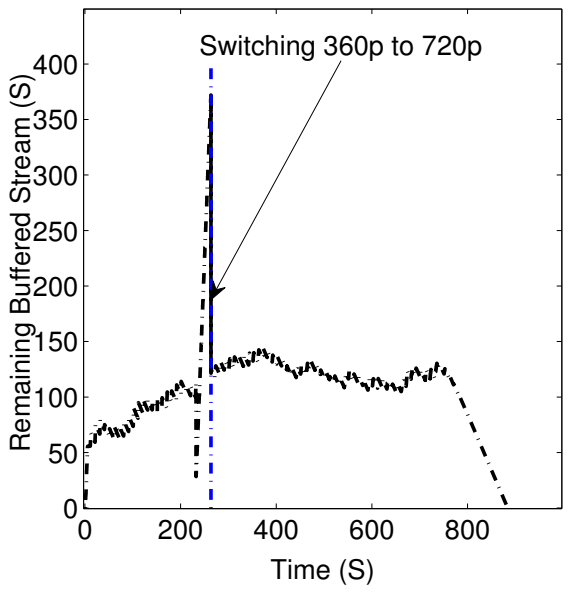

Fig. 18: Playback buffer status during HTTP Live Streaming from Vimeo to iPhone via W-Fi.

## E. Dynamic Adaptive Streaming over HTTP (DASH)

The earlier mentioned streaming techniques allow the client player to watch only one quality of video during a streaming session. Changing quality is possible only by interrupting the playback to switch the video quality. DASH, on the other hand, allows the player to switch the stream quality on the fly in order to adapt to bandwidth fluctuations. Vimeo in iPhone uses Apple's version of DASH called HTTP Live Streaming (HLS) [9].

An HLS server maintains metadata of the video streams describing the quality and required bandwidth between the server and client, such as the following example:

```
#EXT-STREAM-INF:
PROGRAM-ID:1,BW=859000,RES=640x360,index_0_av.m3u8
#EXT-STREAM-INF:
PROGRAM-ID:1,Bw=386000,RES=480x270,index_1_av.m3u8
#EXT-STREAM-INF:
PROGRAM-ID:1,BW=2654000,RES=1280x720,index_2_av.m3u8
```

HLS works so that the player first requests the master playlist file which contains a playlist or index file names for different qualities. Next, the player requests the index file for the default quality video (`index_0_av.m3u8`). This index file contains the list of the segments each of which is then requested and downloaded sequentially. In case of significant bandwidth fluctuations, the player changes the quality and requests higher or lower quality segments.

Figure 18 shows the estimated player buffer status for the whole duration. In the beginning (not shown in traffic graph), the player requests segments periodically every 10 seconds till the 30th segment. At this point the player requests for the highest quality video segments but again from the 24th segment. It downloads from 24 to 34th segments within next 30 seconds. We notice from the buffer status graph that the player constantly maintains a buffer of over 100s.

# V. Streaming Services and Power Consumption

We also measure the smartphone current draw during the streaming sessions. The measured value is for the entire device but we separate this total current drawn into just the average video playback and wireless communication current. Playback includes decoding and display current. We can identify this current draw at the end of the power trace of each streaming session when the content has been fully delivered but playback still continues since some of content is always buffered at the end regardless of streaming technique used. During this time the Wi-Fi or 3G interface is in PSM sleep or CELL_PCH/IDLE mode, respectively. We compute the average wireless communication current, which we refer to as streaming current, by subtracting the average playback current from the average total current.

## A. Impact of Video Quality, Player and Container

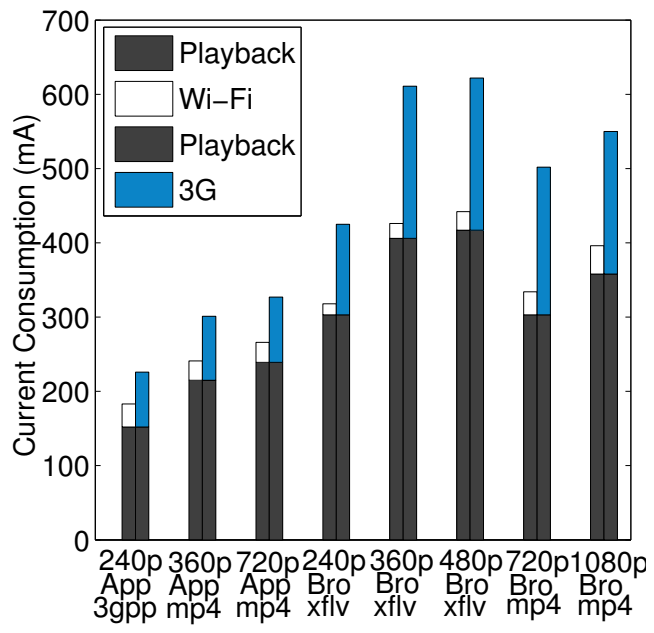

Fig. 19: Avg current draw when streaming $240-1080p$ YouTube videos to the app and browser in Galaxy S3.

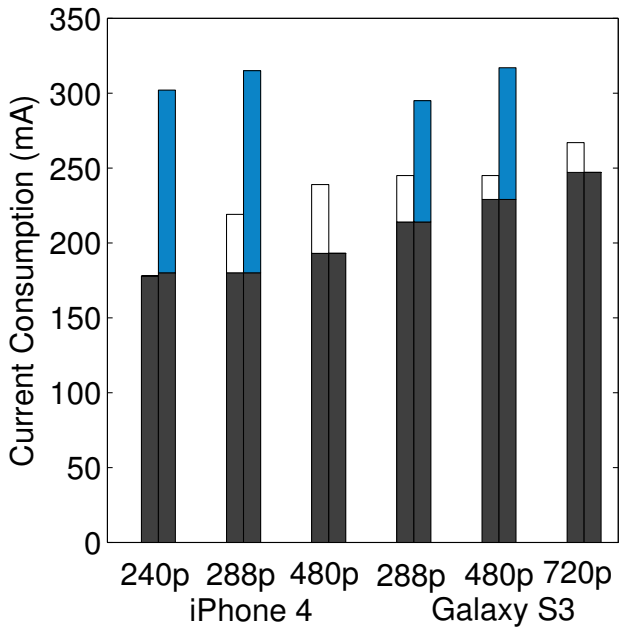

Fig. 20: Avg current draw when streaming $240-720p$ Dailymotion Videos to Galaxy S3 and iPhone 4.

In this section, we discuss how the playback power consumption of smartphones changes with the quality of video, video container, and the player type.

**Video Quality:** From Figure 19, we can see that playback current draw of Galaxy S3 increases as the quality of YouTube video increases using both application and browser. We also observed similar pattern for watching Dailymotion videos in iPhone 4 and Galaxy S3 as shown in Figure 20. It is logical that high quality videos have more information to present than low quality videos and, therefore, more current is drawn. However, in some cases even doubling the resolution adds a relatively small increment to the average playback current.

**Video Player:** In order to play YouTube LD and SD videos, the browser loads Flash player from plug-in. In the case of HD videos, HTML5 player is loaded. We notice that a browser player can draw even a double the current compared to a native app when playing the exact same video.

**Video Container:** We also noticed that playback power consumption differs for video containers. In the case of browser, YouTube server sends LD and SD videos using `xflv` container, whereas the app uses `3gpp` for LD and `mp4` for SD. Figure 19 shows that the average playback current of an `xflv` video is higher than that of the same quality of a `3gpp` or even higher than a HD( $720, 1080p$) quality `mp4` video. We do not currently know the exact cause for these differences between players and containers.

### B. Device Variation and Playback Power Consumption

Figures 21, 22, 23, and 24 compare the average current draw of different devices playing videos from Dailymotion, Vimeo, and YouTube. In the case of Vimeo, iPhone plays SD and HD quality segments from Vimeo and others play only SD ($270p$) quality.

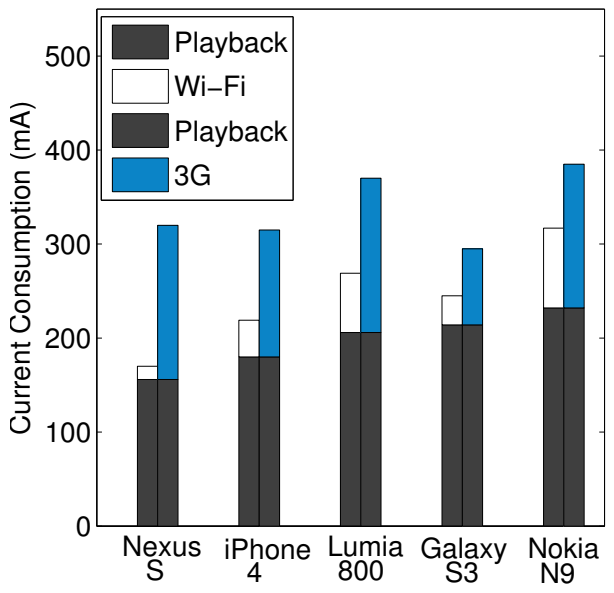

**Fig. 21:** Avg current draw when streaming Dailymotion SD($288p$) video via Wi-Fi and 3G.

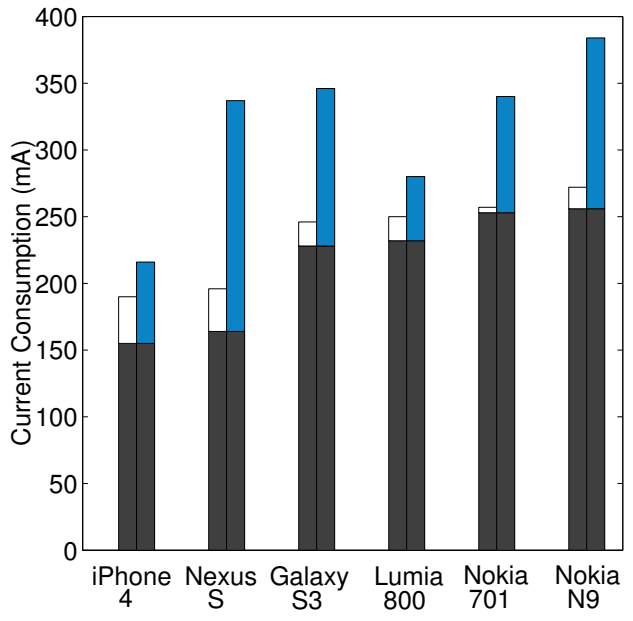

**Fig. 22:** Avg current draw when streaming Vimeo video via Wi-Fi and 3G to smartphones.

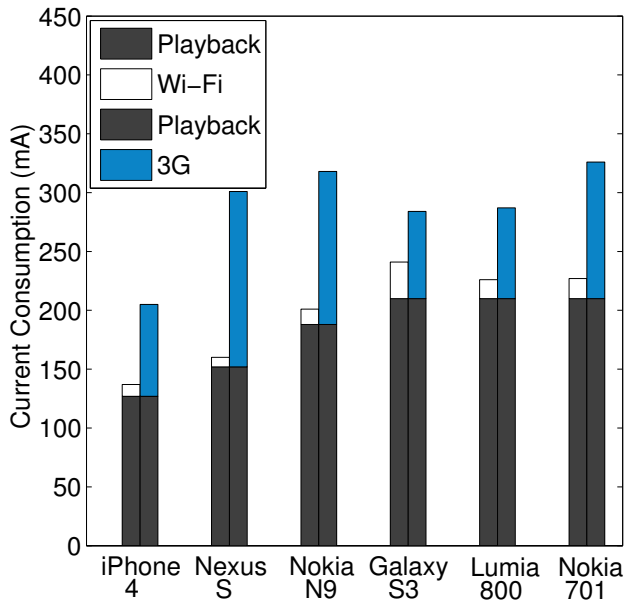

**Fig. 23:** Avg current draw when streaming $240p$ YouTube video using native apps.

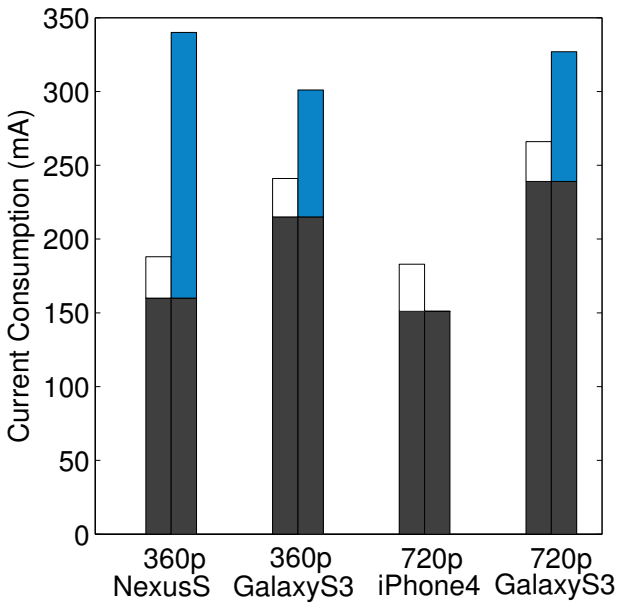

**Fig. 24:** Avg current draw when streaming YouTube SD($360p$) and HD($720p$) video via Wi-Fi and 3G using apps.

Overall, we observe that the average playback current consumption of the same video among multiple smartphones varies significantly, as expected. This variation could be caused by the display resolution, display size, and/or display type. The resolution is unlike the main cause because iPhone 4 has a higher resolution display than any other phone except for Galaxy S3 and at the same time is among the least playback current consuming devices. On the other hand, the size of the iPhone's display is the smallest which reduce current consumption. However, Nokia 701 has an equally small size display and it is among the most playback current consuming devices. Different display technologies are also used including AMOLED, Super AMOLED, IPS LCD. Furthermore, the phones have different GPUs. It looks like the root cause of the differences would be a combination of features and technological differences, but at this point, it is difficult to say anything more conclusive without further experimentation.

### C. Impact of Streaming Techniques

The Figures in the previous sections (V-A and V-B) illustrate that the playback always consumes more current than the wireless communication although in some cases the two equal. Another easy observation is that streaming via 3G always draws more current than streaming over Wi-Fi. In this section we discuss the effect of the different streaming techniques. In the comparison shown in Figure 23, iPhone 4 uses throttling, Nexus S and Galaxy S3 apply On-Off technique, Lumia 800 use Fast Caching, and Nokia 701 and N9 use encoding rate streaming (cf. Table I).

*1) Encoding Rate Streaming:* In this case, the content is delivered continuously throughout the entire streaming session and the wireless interface is all the time active. As a consequence, the average streaming current drawn by Nokia 701 and N9 are among the highest in the figure. Figure 19 shows another example, where Galaxy S3 consumes around 30 mA for Wi-Fi and 200 mA for 3G (HD video using browser). The high current consumption of 3G is expected since the interface is constantly in the high power consuming DCH state. However, power consumption over Wi-Fi is low with respect to the usage of the interface. It could be that the smartphones use some physical layer mechanism where power consumption of the interface is dynamically controlled according to the bit rate [10].

*2) Throttling:* Using this mechanism, the length of the video delivery phase depends on the throttle factor, which in turn determines how long the 3G or Wi-Fi radio will be powered on. Figure 19 includes the throttling cases for Galaxy S3 where the browser is used to stream LD and SD videos from YouTube. In this case throttle factor is 1.25. A similar case is shown in Figure 21 for Nokia N9 streaming from Dailymotion. Both exhibit comparatively high current consumption over 3G, but Galaxy S3 draws clearly less current than N9 when streaming over Wi-Fi. The reason is that YouTube server sends traffic in small chunks, as we explained in Section IV-B, and Wi-Fi interface manages to transition to sleep state in between bursts due to the short timeout ( 100ms). 3G cannot leverage these traffic patterns because the inactivity timer values are much longer than the burst intervals. iPhone 4 consumes even less current in both cases, Wi-Fi and 3G, as in addition to the bursty traffic patterns, the content is downloaded faster, i.e. at twice the encoding rate (see Figure 23).

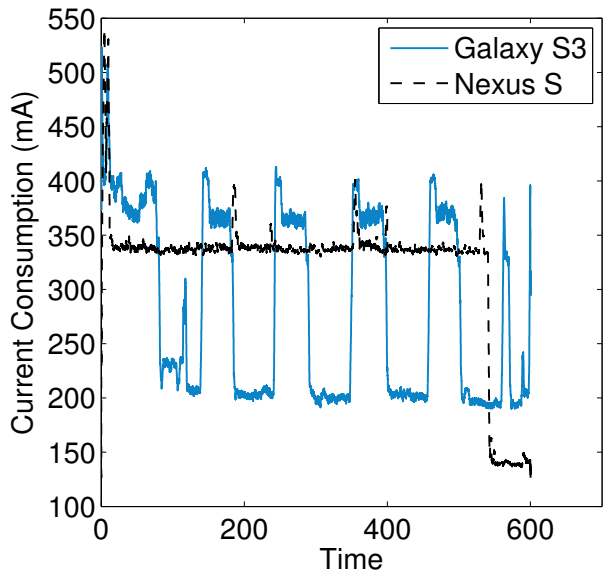

**Fig. 25:** Nexus S is constantly in DCH state whereas Galaxy S3 alternates between DCH and lower power states when streaming using On-Off technique over 3G.

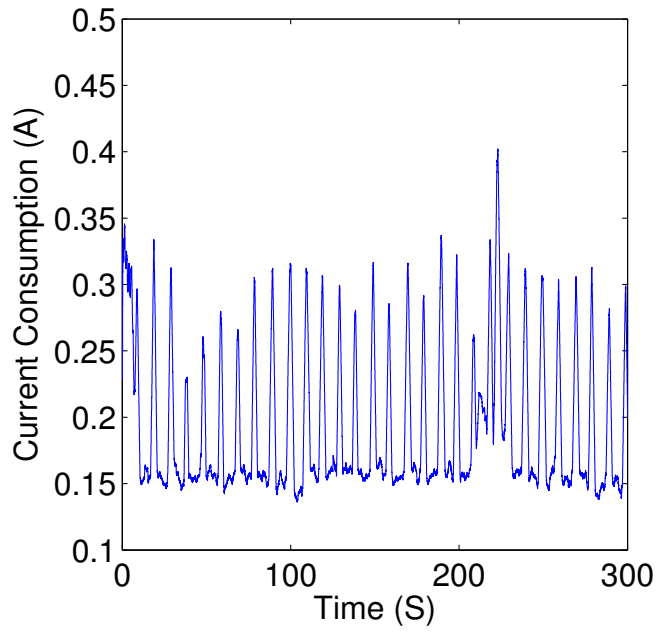

**Fig. 26:** Power trace of watching YouTube on iPhone using DASH over Wi-Fi.

*3) On-Off:* Figure 23 shows that Nexus S consumes very little streaming current over Wi-Fi. In contrast, it uses a lot over 3G. The reason is the use of a single TCP connection which sends the TCP zero win advertisements and probes constantly and keeps the 3G interface in DCH state all the time. We computed from the packet trace that the maximum interval between these packets is 5 seconds, while the 3G's T1 timer is set to 8s. Wi-Fi can sleep most of the intervals in between these control packets. The same figure also includes a case of Galaxy S3 which applies the On-Off technique using multiple TCP connections in between which there are no packets exchanged. Even though in both cases energy is saved over Wi-Fi compared to encoding rate streaming, Galaxy S3 saves more than 50% in average streaming current when streaming over 3G compared to Nexus S. The difference is clearly visible in Figure 25 which compares the power traces from these two cases of streaming over 3G.

*4) Fast Caching:* Fast caching downloads the content with as high throughput as possible. As a result the wireless interface is maximally utilized for as little time as possible. As a consequence, the average current draw is also low. For example, streaming the Dailymotion SD video to Galaxy S3 (Figure 21), Vimeo SD video to Lumia 800 and Nokia 701 (Figure 22), and YouTube LD video to Lumia 800 (Figure 23) consume less streaming current via Wi-Fi and 3G than any other mechanisms. An exception is the case of streaming Vimeo SD video to Nokia 701 (Figure 22) over 3G. In this case we used a browser to play the video and observed unexpected frequent authentication messages between the mobile phone and server. Therefore, the phone alternates between DCH and PCH state frequently even after the video is completely downloaded which happens during the first 40 seconds.

The Vimeo player only in iPhone uses HTTP rate adaptive streaming. For this reason it is difficult to contrast the results from that experiment since the video quality also changes in the middle of the stream. The player receives video content as segments of roughly 10 seconds which enables the wireless interfaces consume less power as the Wi-Fi interface can sleep in between receiving the segments as shown in Figure 26. In Figure 22, we can see that iPhone consumes a little bit more communication power than Lumia 800, even though Vimeo player in iPhone downloads three times more data than Lumia 800.

### D. Discussion

We found out in the previous section that On-Off with separate connections and Fast Caching seem to consume the least energy compared to the streaming at encoding rate and throttling. This result is logical since we know from earlier research that in general the energy efficiency improves with increasing throughput [11], [12]. However, if the user does not watch the whole video, the energy efficiency gets worse because both techniques download significant amount of video data in advance, as we saw in Figures 15 and 17. This energy waste can be significant and happen quite frequently. For example, in [8] Finamore et al. analyzed YouTube traffic to desktop computers and iOS devices accessed via Wi-Fi and discovered that 60% of videos watched for less than 20% of their duration. In Figure 27, we plot the average current draw for example cases of both techniques as a function of percentage of watched video computed out of the complete power traces. We see that interrupting the video watching early on would cause a hefty penalty in terms of wasted energy in both cases but the penalty gets smaller faster with the On-Off streaming making it a more attractive technique since it is common to not watch the video completely.

We have also learned in the previous section that the choice of the network interface has a large impact on the energy consumption, as could be expected. An interesting observation is that on several of the smartphones, such as the N9 and 701, the current draw of Wi-Fi is quite insensitive to the streaming technique used. This result suggests that perhaps these WiFi modules use, for instance, some sort of dynamic modulation scaling in order to achieve such power proportionality. On the other hand, the streaming technique matters a lot when using 3G. In light of these findings, we find it rather surprising that none of the services or client players varied the streaming technique depending on the access network type used, which leaves clearly room for optimization.

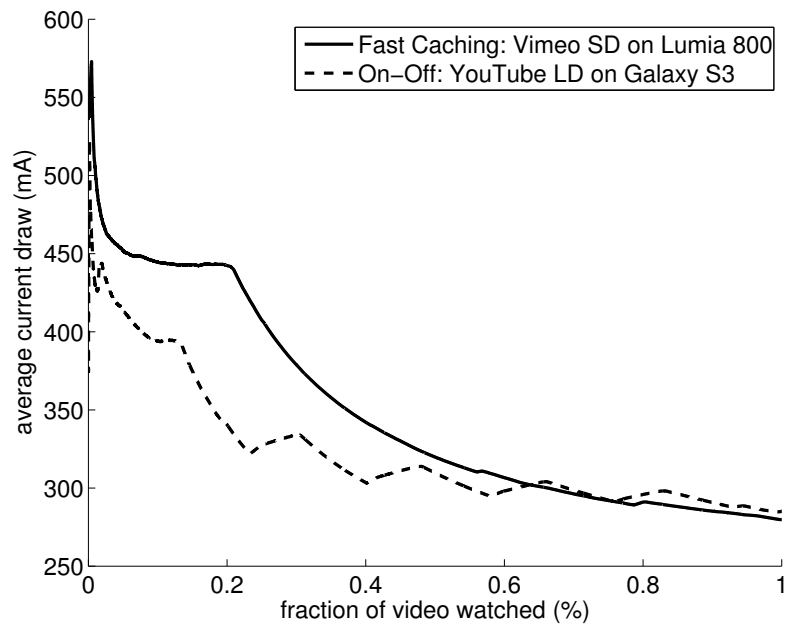

**Fig. 27:** Average draw of current as a function of fraction of watched video over 3G.

We obtained our results in close to "ideal" conditions in that we used a private 3G access in isolated RF room in order to get comparable results. However, we also did a round of measurements for most of the services and devices where we limited the available bandwidth to close to the video encoding rate using a software based rate limiter. In this way we emulated a case of a loaded cell in 3G network. The main observation was that the different techniques start to resemble encoding rate streaming when comparing traffic patterns and resulting energy consumption, which is an expected result since there is no longer left over bandwidth to leverage. Signal strength can also vary depending on the user's location within a cell. We estimate that in such a case all the techniques would be penalized by having to use more transmit power and to amplify more the received signal. Such situation would hurt more the techniques that need to keep radio in rx/tx mode longer time meaning that the difference in average current drawn between the most and least energy efficient techniques would increase.

## VI. Related Work

The diverse nature of existing popular mobile streaming services and the resulting energy consumption characteristics have so far not been completely uncovered. Many papers have studied the energy efficiency of multimedia streaming over Wi-Fi and developed custom protocols or scheduling mechanisms to optimize the behavior. Examples of such work range from proxy based traffic shaping [13] and scheduling [14] to traffic prediction [15] and adaptive buffer management [16]. However, streaming over 3G and the specific nature of the streaming services and client apps provide new challenges that these solutions cannot overcome. Balasubramanian et al. studied 3G power characteristics in general and quantified the so called tail energy concept [12]. However, their work did not consider streaming applications. Even the use of Bluetooth has been suggested to save energy while streaming [17].

The most popular streaming services, especially YouTube, have been subject to numerous measurement studies in recent few years. In [18], authors discovered that YouTube traffic pattern is bursty which makes it by default energy-efficient and they proposed a client-controlled solution leveraging TCP flow control to increase the burstiness and, hence, energy efficiency. Unfortunately, this clever solution suffers from the same problem as persistent connection On-Off patterns (Figure 14), i.e. the TCP zero window probe and advertisement messages keep the network interface powered up, especially in the case of 3G.

Xiao et al. [19] measured the energy consumption of different Symbian based Nokia devices while using a YouTube application over both Wi-Fi and 3G access. That study did not consider the details of YouTube traffic patterns and their impact on the energy consumption.

In a measurement study, Rao *et al.* [7] studied YouTube and Netflix traffic to different smartphones (iOS and Android) and web browsers accessed via Wi-Fi interface. They found three different traffic patterns of YouTube. In a similar passive measurement study [8], Finamore et al. also analyzed YouTube traffic to PCs and iOS devices accessed via Wi-Fi and demonstrated that iPhone and iPad employ chunk based streaming. In contrast to these studies, we investigated which characteristics influence the choice of the streaming technique and quantified their impact on the energy consumption on different smartphones including also 3G access into the study.

Qian *et al.* explored RRC state machine settings in terms of inactivity timers using real network traces from different operators [20], [21] and proposed a traffic shaping solution for YouTube streaming called chunk-mode. Their solution closely resembles the on-off streaming technique we identified. Their study also suggests using dynamic settings of the inactivity timers based on observed traffic patterns, which would require changes to the software controlling the state transitions at the base station.

Liu et al. studied power consumption of different streaming services [22]. However, the scope of their study is considerably different from ours. They limit their study to streaming over Wi-Fi and performed experiments with only iPod, while we explored all the major mobile platforms and contrasted Wi-Fi with 3G. Our methodology is centered around fine grained power measurements with external instruments in controlled environment, while Liu et al. relied on power consumption estimates. They considered also RTSP and P2P streaming which we did not find being in use for the most popular services and mobile platforms.

## VII. Conclusions

We analyzed the performance, and especially the energy consumption, of mobile video streaming. Based on measurements with six smartphones and three popular services we identified five different streaming techniques. The used technique seems to depend on the service, client device, player type, and video quality. Interestingly, the radio interface (WiFi vs. 3G) does not influence the used technique.

Based on our results, the most energy efficient techniques are fast caching and on-off streaming. However, especially fast caching causes unnecessary data transfers because users often watch only part of the video. From this perspective on-off streaming provides a better trade-off. In addition, the video codec as well as the container for the content have a significant impact on the energy consumption and the energy consumption between devices differ a lot due to differences in HW and SW components.

As future work, we intend to study LTE to complement 3G and Wi-Fi results. Furthermore, the impact of Continuous Packet Connectivity (CPC) in 3G and DRX in LTE are interesting subjects to study. Another topic is to quantify the impact of noisy and congested network conditions to the performance of the video streaming services.

In general, we can say that in most cases the video streaming energy consumption is far from optimal. Video streaming service providers, device vendors and network operators can impact significantly the battery life time by selecting optimal video player, coding and container solution, by delivering the content over the radio with the most energy efficient traffic model, and by optimizing the radio parameters accordingly.